# Why most studies into COVID19 risk factors may be producing flawed conclusions - and how to fix the problem

Norman Fenton[1]

Queen Mary University and Agena Ltd

17 May 2020

THIS IS A PREPRINT OF A PAPER INTENDED TO BE REVISED AND SUBMITTED FOR PUBLICATION

## Abstract

An important recent preprint by Griffith et al highlights how 'collider bias' in studies of COVID19 undermines our understanding of the disease risk and severity. This is typically caused by the data being restricted to people who have undergone COVID19 testing, among whom healthcare workers are overrepresented. For example, collider bias caused by smokers being underrepresented in the dataset *may* (at least partly) explain empirical results that suggest smoking reduces the risk of COVID19. We extend the work of Griffith et al making more explicit use of graphical causal models to interpret observed data. We show that their smoking example can be clarified and improved using Bayesian network models with realistic data and assumptions. We show that there is an even more fundamental problem for risk factors like 'stress' which, unlike smoking, is more rather than less prevalent among healthcare workers; in this case, because of a combination of collider bias from the biased dataset and the fact that 'healthcare worker' is a confounding variable, it is likely that studies will wrongly conclude that stress *reduces* rather than *increases* the risk of COVID19. Indeed, the same large study showing the lower incidence of COVID among smokers also showed a lower incidence for people with hypertension, and in theory it could also lead to bizarre conclusions like "being in close contact with COVID19 people" reduces the risk of COVID19. To avoid such potentially erroneous conclusions, any analysis of observational data must take account of the underlying causal structure including colliders and confounders. If analysts fail to do this explicitly then any conclusions they make about the effect of specific risk factors on COVID19 are likely to be flawed.

[1] n.fenton@qmul.ac.uk

# 1. Introduction

A very interesting preprint (Griffith et al., 2020) highlights how 'collider bias' in statistical studies of COVID19 – such as the large-scale study (Collaborative et al., 2020) – may undermine our understanding of the disease risk. The bias is introduced by relying on datasets of patients who have been tested for COVID19 and this testing has not been random; for example, in the UK up until the end of April 2020 this data consisted almost exclusively of patients hospitalized with severe COVID9 symptoms. Since then routine testing has been extended to healthcare workers, but this introduces a new bias.

The Griffith et al paper is excellent as it also illustrates implicitly the need to consider causal graphical models (Pearl & Mackenzie, 2018) when interpreting and analysing observed data. One of the paper's main examples highlights how collider bias may explain recent empirical results which claim that smoking reduces the risk of COVID19. However, their example is somewhat misleading and can be clarified using Bayesian network (BN) models - which are causal graphical models that also incorporate the probabilistic relationships between the variables (N. E. Fenton & Neil, 2018; Pearl, 1988). Using such models, we are able to show that, because of a combination of collider bias and confounding variables, any conclusions about the effect of specific risk factors on COVID19 are likely to be flawed if they are based only on data from people who have been tested. However, in the absence of genuinely random testing, we can avoid the problem using the same (biased) data providing that we adjust the analysis to take full account of the underlying causal structure.

The paper is structured as follows: We provide a brief overview of collider and confounder variables in Section 2. In Section 3 we present the smoking example and show how the (Griffith et al., 2020) partial explanation of it can be formally represented as a simple BN model. The full and more realistic explanation is provided in Section 4. In Section 5 we provide the more worrying example of how conclusions about risk factors can be especially compromised by a combination of confounder and collider variables. The conclusions and way forward are presented in Section 6.

## 2. Colliders and confounders

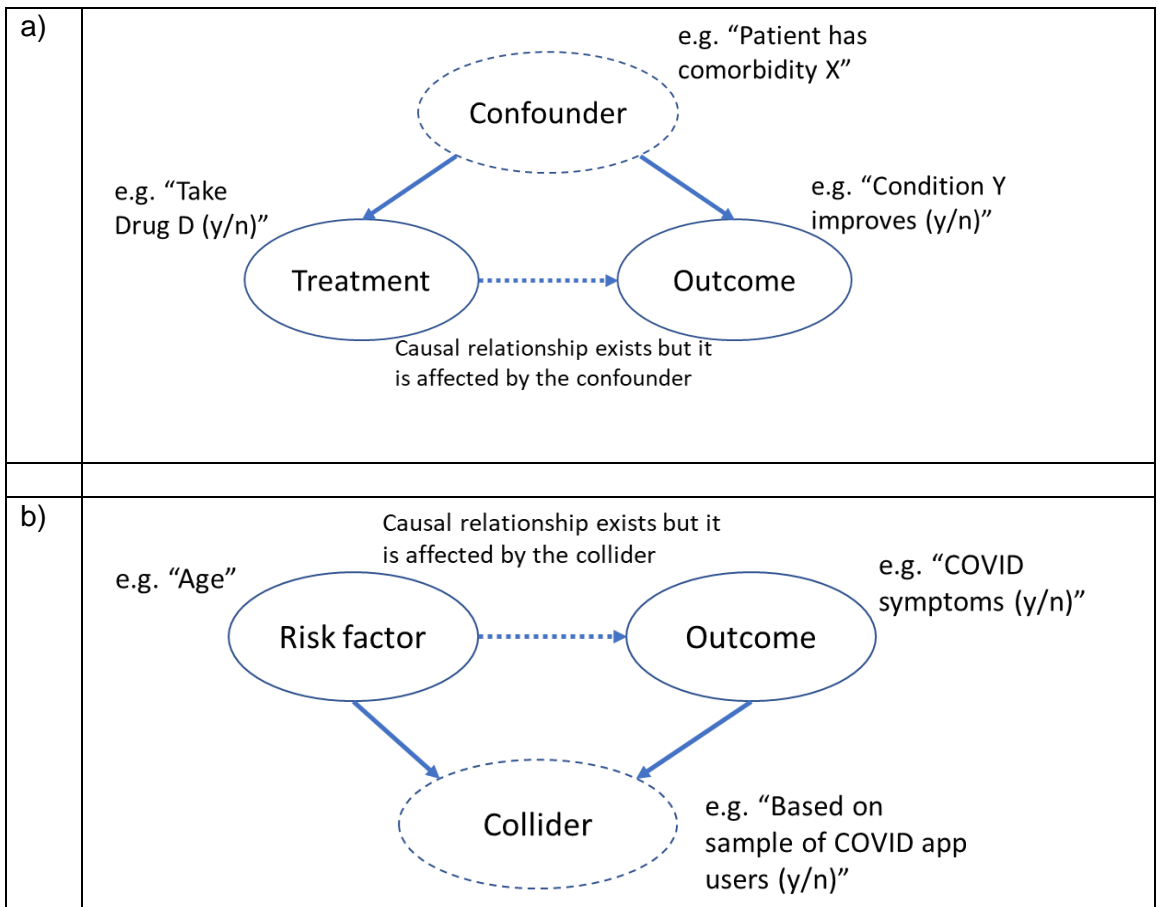

Figure 1 Confounders (a) and Colliders (b)

As shown in Figure 1(a), a confounding variable is one which is a common influence on two variables. For example, whether patients with condition Y have comorbidity X may influence both whether they are given drug D and whether their condition Y improves. In contrast, as shown in Figure 1(b), a collider is a variable that is influenced by two other variables of interest. For example, if you are investigating the relationship between a risk factor like age on whether a person gets COVD19 symptoms by using data from a COVID phone app, then it is important to note that young people and the 'worried well' are more likely to use the app than very old people with symptoms.

If not taken into consideration in observational studies, the presence of confounders and colliders can introduce bias and lead to false conclusions about the relationship between those two variables. For example, the presence of a confounder may lead to *Simpson's paradox* (N. Fenton, Neil, & Constantinou, 2019), whereby the treatment may appear to have a positive effect for the population of patients overall, yet have a negative effect for each relevant subclass of the population (such as patients with the comorbidity and patients without the comorbidity). The presence of a collider may lead to *Berkson's paradox,* whereby a risk factor that has no effect or a negative effect on the outcome is shown to have a positive effect when the data are restricted to the collider; a classic non-medical example is shown in Figure 2 based on an example from (Pearl & Mackenzie, 2018). In this example 'mean' people are under-represented in the dataset used for analysis, and as a result attractive people are perceived to be more likely to be mean (or equivalently less likely to be nice).

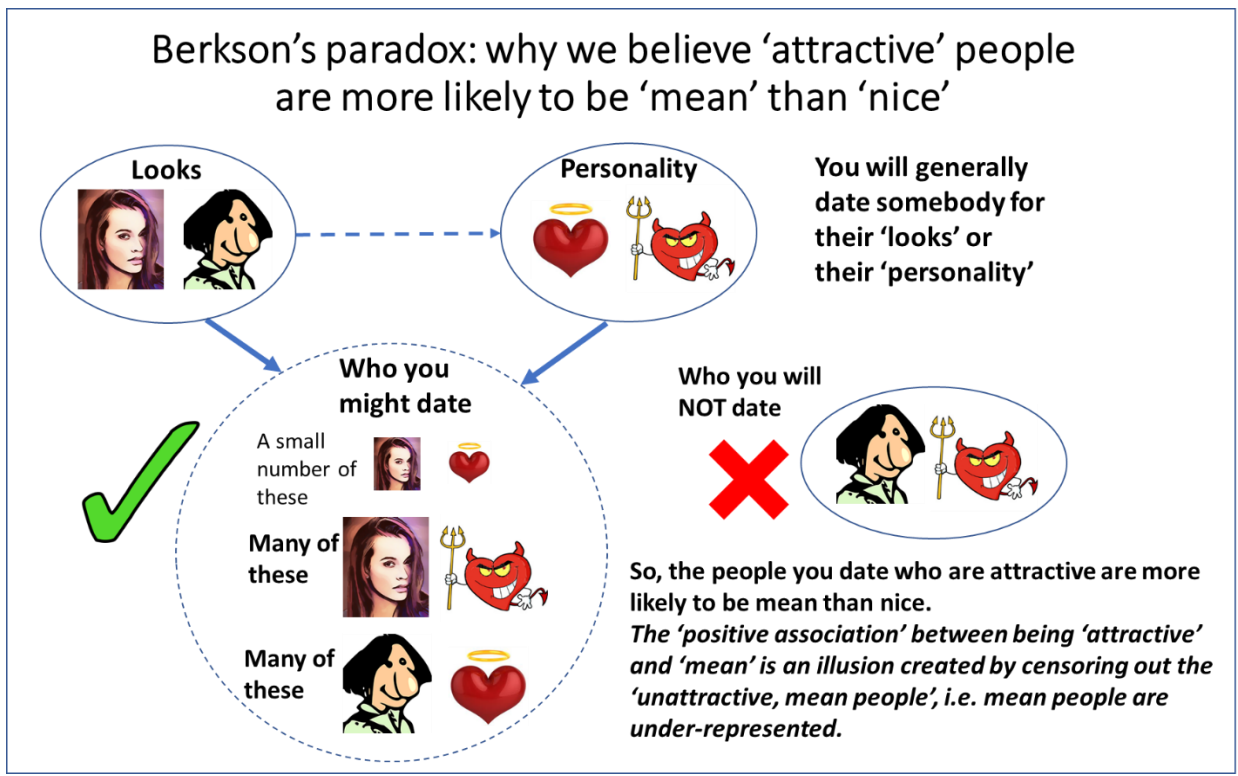

*Figure 2 Berkson's paradox: Basing the study on 'who you might date' introduces a collider*

## 3. The (over-simplified) smoking example using a causal Bayesian network

(Griffith et al., 2020) argue that many of the current COVID-19 datasets rely on non-random participation with strong selection pressures. Most notably, many of the datasets are based on people who have been tested for COVID-19, and these are dominated (and hence biased) by two groups of people:

- People already hospitalized with severe COVID-19 symptoms
- Healthcare workers

Indeed, in the UK at time of writing these are still the only people routinely receiving tests.

So, while studies show apparently counterintuitive results such as that smoking appears to reduces the risk of COVID-19 symptoms and death (Collaborative et al., 2020; Miyara et al., 2020), Griffith et al argue that collider bias may explain such results; this is in contrast to (Changeux, Amoura, Rey, & Miyara, 2020) who argue that nicotine may have preventive and therapeutic value in this context.

However, there are problems with the way the smoking example is presented in (Griffith et al., 2020). In particular, their graphical model – reproduced here in Figure 3 is confusing and is actually not used in the paper to explain how collider bias can lead to flawed conclusions about the benefits of smoking. In fact, (Griffith et al., 2020) implicitly assume a simpler model to explain this.

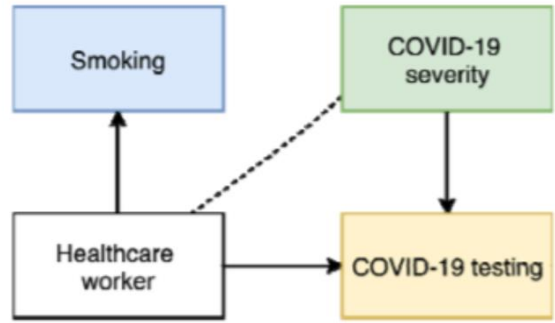

*Figure 3 "Sampling conditioned on testing" - reproduced from (Griffith et al 2020)*

The Griffith et al argument is essentially that shown in Figure 4, where we have adapted the previous Berkson paradox example replacing 'looks' with 'smoking' and 'personality' with 'COVID19'.

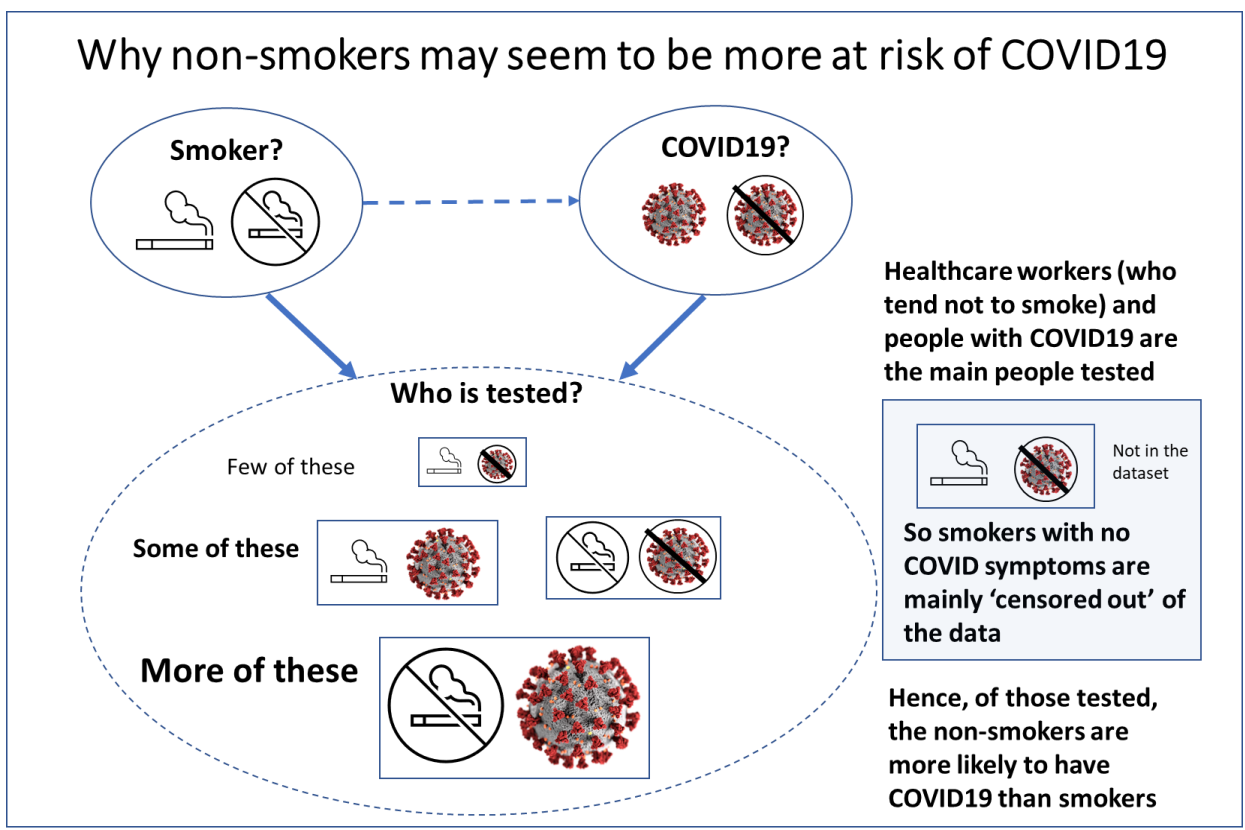

*Figure 4 Berkson paradox argument explaining why non-smokers may be (erroneously) seen to be more at risk of COVID19*

Using the same assumptions as in (Griffith et al., 2020)), we first use the Bayesian network (BN) shown in Figure 5 to formalise how Griffith et al actually explain the collider problem. Note that a BN consists of both the graphical model and associated probability tables for each node.

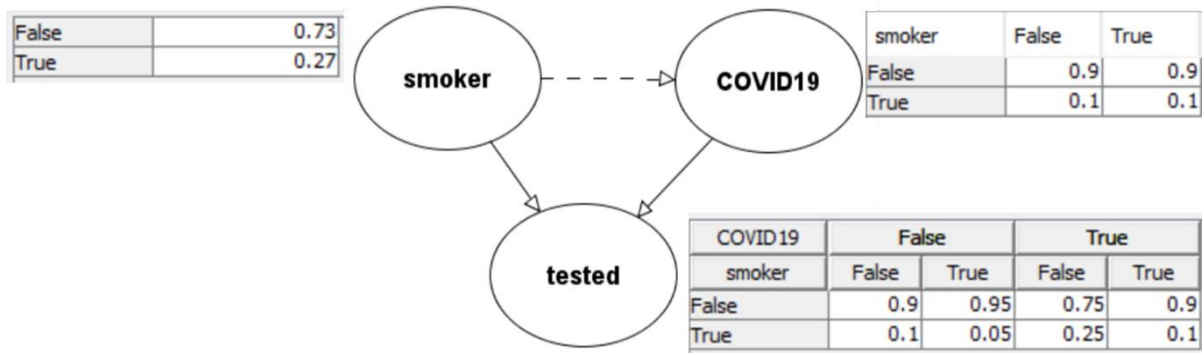

*Figure 5 Simple BN structure and prob representing collider model for effect of smoking*

The starting assumption is the 'null hypothesis' that smoking has no impact on the risk of COVID19 (hence the dashed rather than solid line in Figure 5), The objective is to show that the collider effect can produce (erroneous) 'evidence' that smoking reduces the risk. The following assumptions are encoded into the probability tables associated with the model:

- 27% of the general population are smokers. This explains the probability table associated with node 'smoker'.
- 10% of the population have COVID19 (we could use any percentage and it does not affect the argument), with no difference between smokers and non-smokers (the latter is just the null hypothesis). This explains the (conditional) probability tables associated with node COVID19;
- Smokers are under-represented among those tested. This is the crucial assumption made, because we are restricting the analysis to those who are tested. As we saw in Figure 4 smokers are under-represented in those tested leading to the wrong conclusion that smokers are less likely to have COVID19. We use the following explicit probabilities, which explain the conditional probability table associated with the node 'tested' (we could have used any values which fall into the ranges specified by (Griffith et al., 2020)):
    - 10% of smokers with COVID19 – compared to 25% of non-smokers with COVID19 – are tested.
    - 5% of smokers without COVID19 – compared to 10% of non-smokers without COVID19 – are tested.

Figure 6(i) shows the results of running the model[2] without any observations, i.e. without the assumption the we are restricting the analysis to those who have been tested (so, the probabilities shown for 'tested' is the result of the standard formula for computing marginal probabilities from conditional probabilities). Figure(ii) shows the result of running the model **with** the assumption that we are restricting the analysis to those who have been tested (this is simply applying Bayes' theorem). Note how this shows that smokers are under-represented, since only 15% are smokers, compared to 27% in the general population.

---

[2] While all the calculations could be done manually using Bayes theorem, the models are instead run in AgenaRisk using its standard Bayesian network inference algorithm. All of the models are in the downloadable file http://www.eecs.qmul.ac.uk/~norman/Models/colliders_paper.cmpx which can be opened and run using the free trial version of AgenaRisk https://www.agenarisk.com/

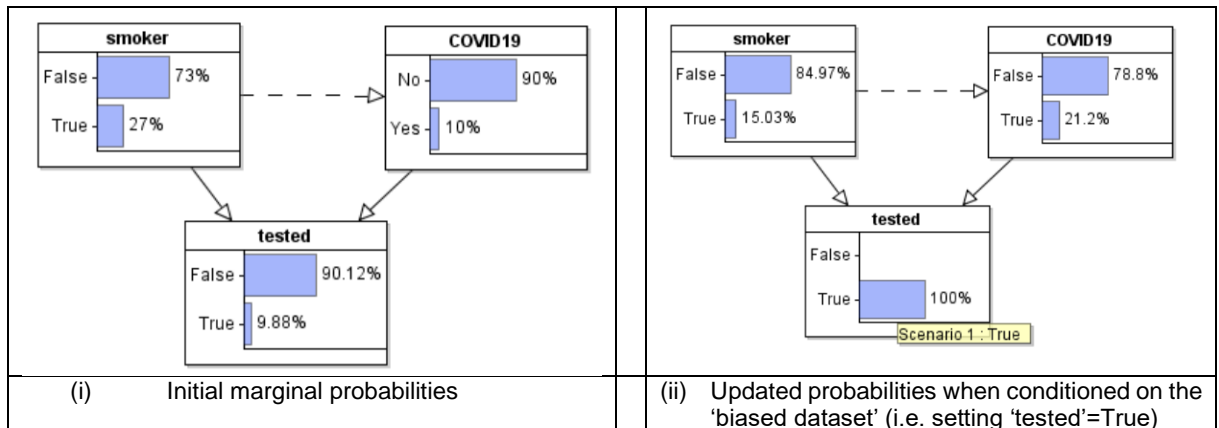

(i) Initial marginal probabilities

(ii) Updated probabilities when conditioned on the 'biased dataset' (i.e. setting 'tested'=True)

*Figure 6 Probabilities before and after conditioning on the collider (i.e. 'tested'=True*

Figure 7 shows the results of running the model (comparing smokers and non-smokers) when we condition on those tested. Smokers are less likely (18.1%) to have COVID19 than non-smokers (21.7%).

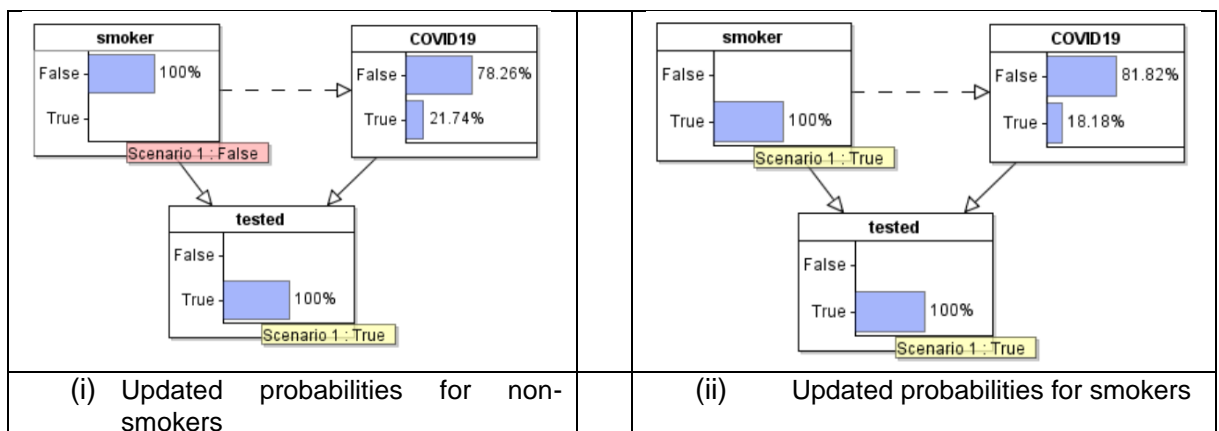

(i) Updated probabilities for non-smokers

(ii) Updated probabilities for smokers

*Figure 7 Updated probabilities when we DO condition on those tested*

So, the example demonstrates that under the hypothesis that smoking has no impact on COVID19 in the general population, when restricted to people who have been tested, smokers have a lower risk than non-smokers. Hence, we have shown formally how the testing bias collider can lead to flawed conclusions about the impact of risk factors.

In fact, with a very small change, we can simulate the even more dramatic situation whereby **it can be wrongly shown that smoking leads to a reduced risk of COVID19 if we consider only people tested** under the hypothesis that **smoking leads to greater risk of COVID19 in the general population.** Suppose for example that, in contrast to the previous null hypothesis, smoking is known to increase the probability of COVID19 among smokers from 10% to 11%. Then making only this change to the conditional probability table for COVID we still get the 'reversal' when we condition on those tested. This is shown in Figure 8.

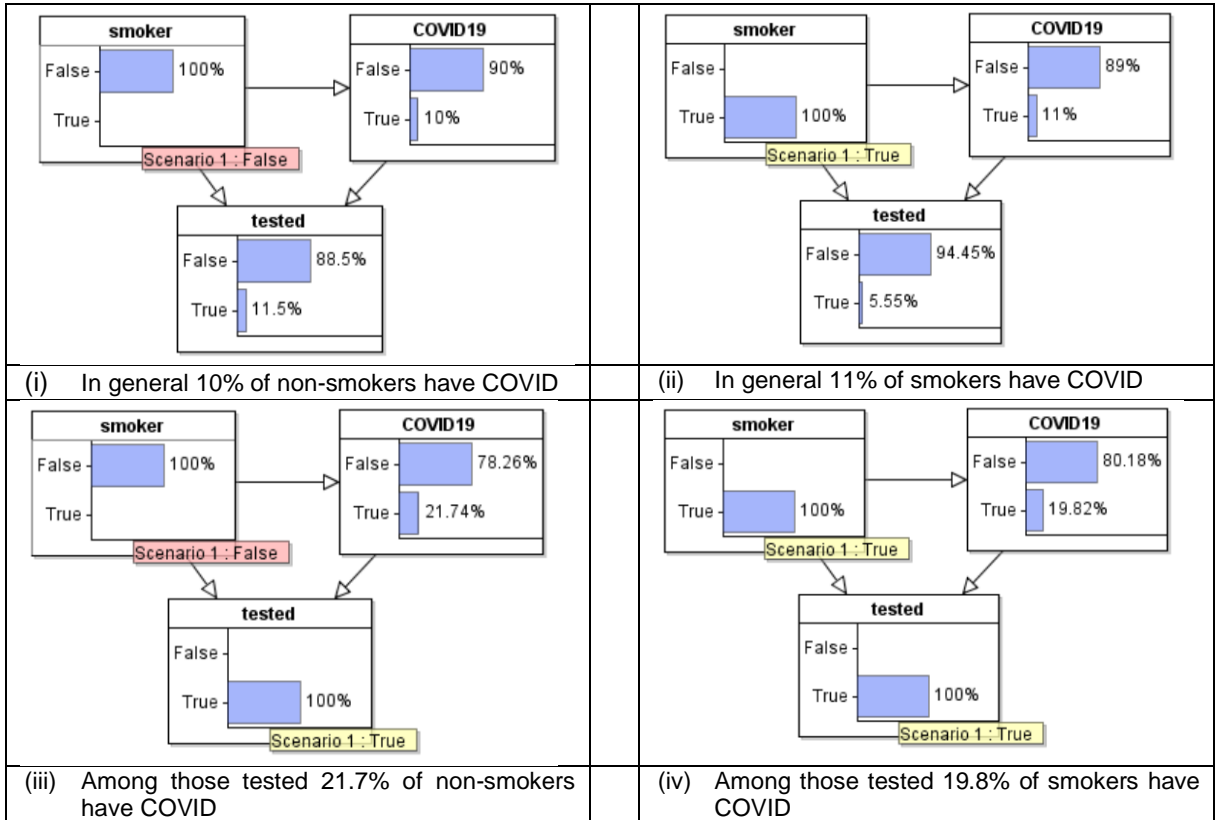

Figure 8 Model in which smokers are more at risk in the general population (as shown in (i) and (ii)). Yet when restricted to those tested smokers are less at risk (iii) and (iv)

## 4. A more realistic smoking example

While the above 3-node BN model (like the narrative in Griffith et al) explains how a bias in those tested can, in theory, lead to flawed conclusions what it fails to do is provide an explanation of how this can feasibly happen in practice. However, by using a causal BN model that 'fixes' the problem with the Griffith et al model in Figure 3, we can do exactly that. The model required is shown in Figure 9, along with the prior and conditional probability table assumptions.

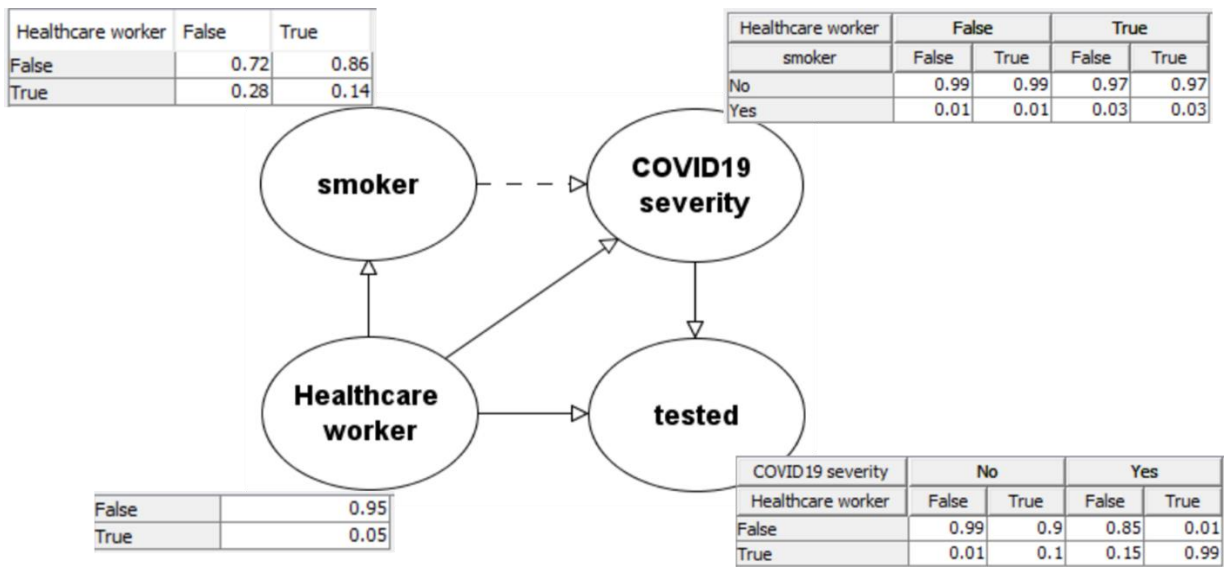

Figure 9 Causal BN representing collider model for effect of smoking

The objective is to show (as before) that, under the 'null hypothesis'[3] that smoking does not impact the risk of COVID19, the collider effect can produce 'evidence' that smoking reduces the risk. First note that the probability table parameters mean that we assume:

- Healthcare workers are more likely to get severe COVID19 than general members of the public. We assume 3% probability of severe COVID19 for healthcare workers compared to 1% for general members of the public based on commonly cited figures[4]. As we are assuming the null hypothesis that smoking does not impact the risk of severe COVID19, the conditional probability table for COVID19 severity in Figure 8 is the same for smokers and non-smokers.
- 5% of the population are healthcare workers[5] and that healthcare workers are less likely to be smokers than general members of the public (14% compared to 28%).
- Only healthcare workers and people with severe symptoms are likely to be tested. A healthcare worker with severe symptoms is almost certain to be tested (99%, compared to 15% for non-healthcare workers), while a non-healthcare worker without symptoms is almost certain not to be tested (1%, compared to 10% for healthcare workers without symptoms). These assumptions fairly reflect the reality of UK data between early March to late April when many of the studies were conducted.

Running this BN model gives the marginal probabilities shown in Figure 10(i). However, crucially, note how the marginals change in Figure 10(ii) when we condition the data on those tested, i.e. we set the value of 'tested' to be True. This is where the collider bias is introduced - the study has a disproportionately high number of healthcare workers and lower number of smokers than in the general population. And the overall number with severe COVID19 is also disproportionately high.

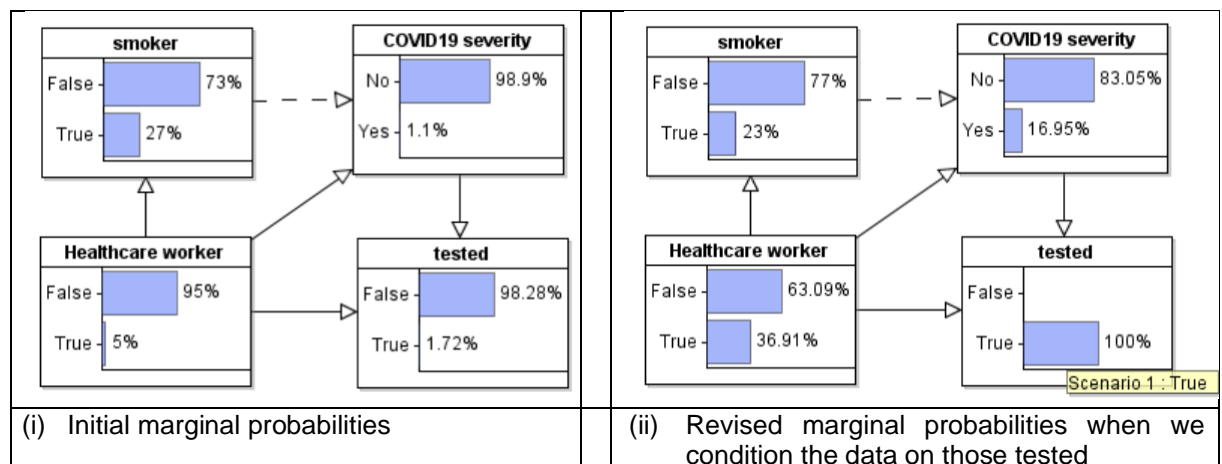

| (i) Initial marginal probabilities | (ii) Revised marginal probabilities when we condition the data on those tested |

*Figure 10 Marginal probabilities before and after conditioning on the data for those tested (i.e. setting 'tested'=true)*

So now when we run the model comparing smokers and non-smokers, as shown in Figure 11, we get the result that smokers have a reduced risk of severe COVID19 (down from 17% to 15%).

---

[3] So again because of the 'null hypothesis' we use a dotted rather than solid line from smoker to COVID
[4] https://www.theverge.com/2020/3/5/21166088/coronavirus-covid-19-protection-doctors-nurses-health-workers-risk
[5] https://fullfact.org/health/how-many-nhs-employees-are-there/

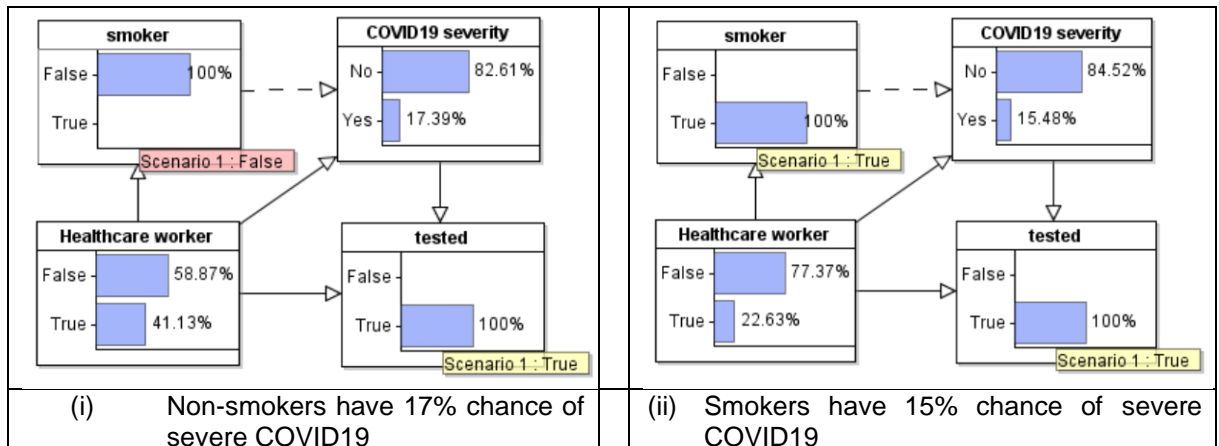

| (i) Non-smokers have 17% chance of severe COVID19 | (ii) Smokers have 15% chance of severe COVID19 |

*Figure 11 Smoking has a beneficial effect in this 'biased' dataset*

So, this demonstrates with a more realistic example how the testing bias collider can lead to a flawed conclusion about the impact of smoking. Moreover, once again, we can easily simulate the even more dramatic situation whereby even if smoking **genuinely leads to greater risk of COVID19, it can be wrongly shown to lead to a reduced risk** because of the collider bias. In fact, if we change the probability table of the COVID19 node by assuming that there is a relative 2% increased risk from smoking (so for non-healthcare workers who are smokers the probability increases from 1% to 1.02% and for healthcare workers who are smokers the probability increases from 3% to 3.06%) then with all other assumptions unchanged we get the 'reversal' when we condition on those tested.

The errors being introduced by testing bias in this example are not solely due to collider bias. In fact, we also have a *confounding variable* - healthcare worker - on COVID19. In this case the effect is subtle because the increased probability of COVID19 for healthcare workers is partly balanced out by the reduced probability of smoking for healthcare workers.

## 5. When confounding variables combined with colliders becomes even more problematic for evaluating risk factors

The confounding effect will be greater in examples where, unlike smoking, the risk factor has an *increased* probability among healthcare workers. One such possible risk factor is 'stress'; indeed the argument below may explain why 'hypertension' was found, like smoking, to reduce the risk of COVID19 in (Collaborative et al., 2020). In fact, the argument presented below will be relevant to every risk factor that is more prevalent among healthcare workers than general member of the public; it could even be used to demonstrate such bizarre results as "close contact with COVID19 people" appears to reduce the risk of COVID19. The relevant causal model is shown in Figure 12.

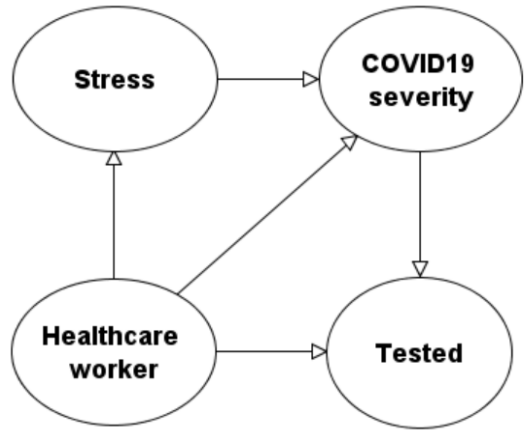

*Figure 12 Causal model for effect of stress on COVID19 where the data is conditioned on those tested and healthcare worker is a confounding variable*

Under the reasonable hypothesis that stress increases the probability of severe COVID19, the combined effect of the testing collider and the healthcare confounder are such that the data would lead to the wrong conclusion that stress **reduces** the probability of severe COVID19. This is shown in Figure 13, which uses the same assumptions about testing as the smoking example but assumes (in contrast to smoking) that healthcare workers are more likely to suffer stress. When we use only the data for those tested, we conclude that people without stress have a significantly higher probability of severe COVID19 than those with stress (15.9% compared to 10.6%).

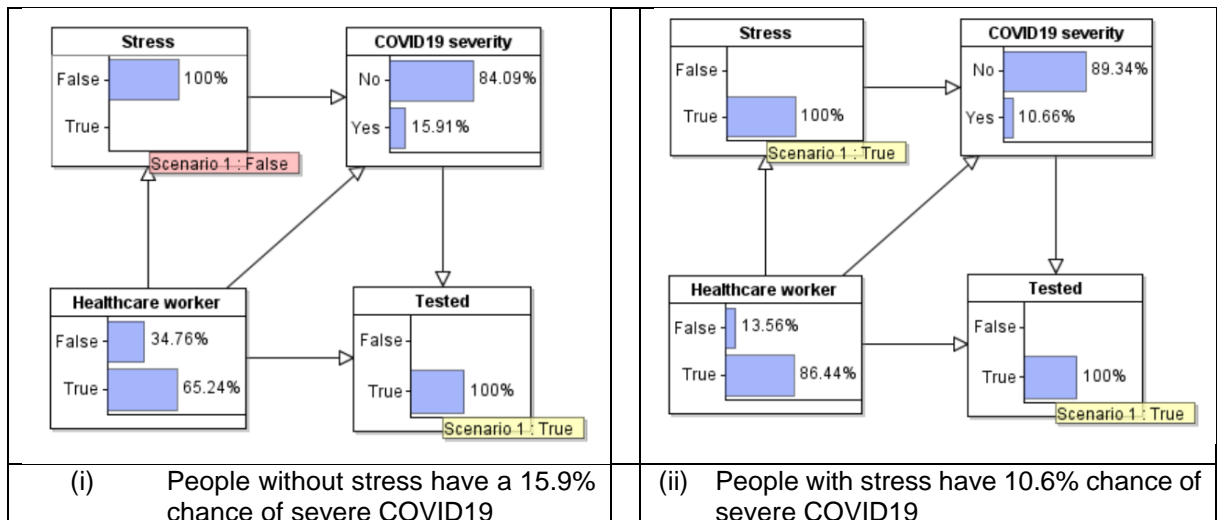

| (i) People without stress have a 15.9% chance of severe COVID19 | (ii) People with stress have 10.6% chance of severe COVID19 |

*Figure 13 Stress has a beneficial effect on this 'biased' and confounded dataset*

To avoid this error, and to determine the 'correct' effect of stress in the general population, not only do we need to remove the 'True' condition on the Tested node, but we also have to break the link from Healthcare worker to Stress in order to simulate the effect of being able to set stress to be true or false (i.e. simulate an intervention). This is shown in Figure 14.

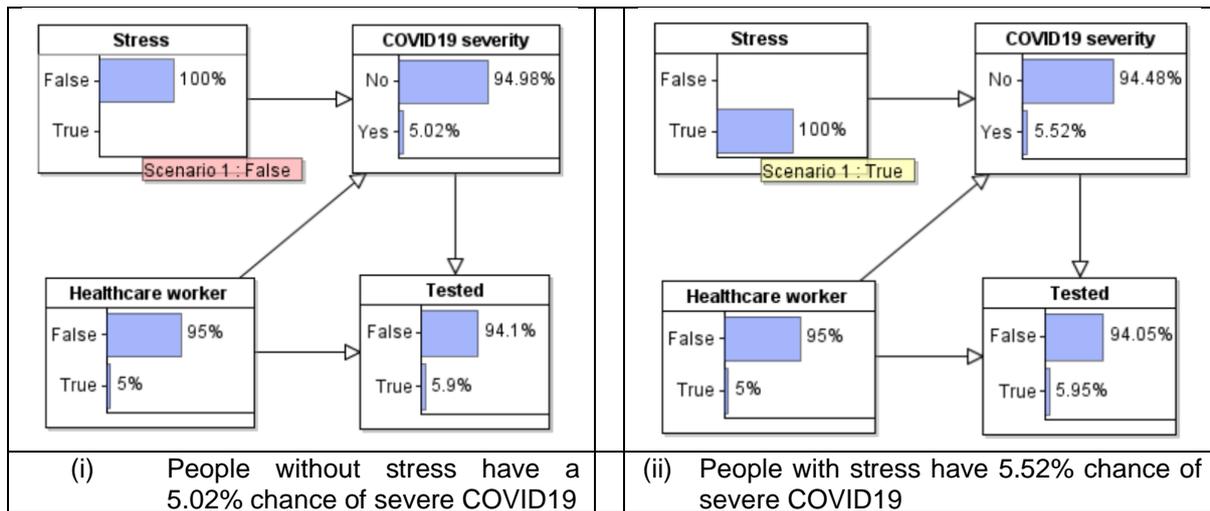

| (i) People without stress have a 5.02% chance of severe COVID19 | (ii) People with stress have 5.52% chance of severe COVID19 |

*Figure 14 To determine the 'true' effect of stress in the general population we have to not only remove the conditioning on those tested, but also break the link from Healthcare worker to Stress since the former is a confounder.*

In the model shown in Figure 15 we replace the risk factor "stress" with "contact with COVID19 patients". Using the reasonable conditional probability assumptions shown in Figure 14, we end up with the bizarre conclusion that, when conditioned on those tested, close contact with COVID19 patients reduces the risk of COVID19 from 7.9% to 6.6%. This is even though the model assumes that, in the general population, close contact with COVID19 patients approximately doubles the risk of COVID19.

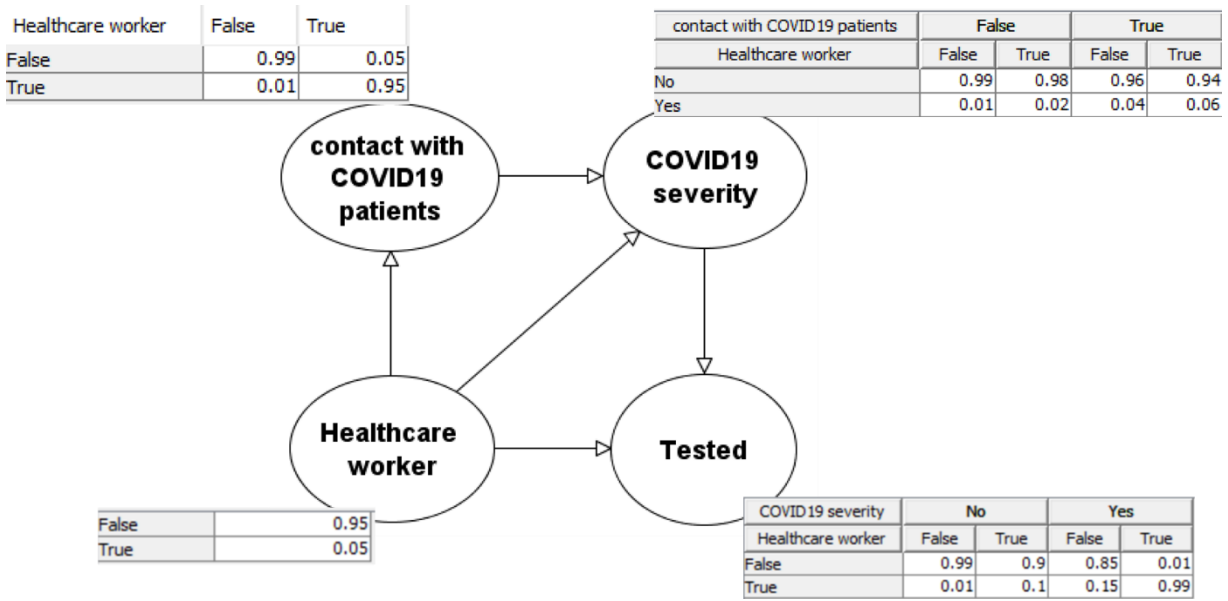

*Figure 15 Model investigating the risk of 'contact with COVID19 patients'*

## 6. Conclusions and way forward

There is a growing awareness – much thanks to the increasingly well publicized work of Pearl (Pearl & Mackenzie, 2018) – of the importance of causal graphical models for data analysis. We have already highlighted the importance of such models for the analysis of COVID19 data for interpreting death rates (N. E. Fenton, Neil, Osman, & McLachlan, 2020), incorporating testing errors (N. Fenton, Hitman, Neil, Osman, & McLachlan, 2020) and symptom tracking (McLachlan et al., 2020). Most studies of COVID risk factors use data for people tested and, because there has been no systematic random testing, this means the data are biased. The important recent work of (Griffith et al., 2020) that was the motivation for this paper illustrates that failing to take account of causal notions like colliders for such biased data, may lead to flawed conclusions about the effect of risk factors on COVID19. By using Bayesian network (BN) models explicitly to model the causal structure and probabilistic strength of relationships between relevant variables, we have shown formally how such flawed conclusions may arise. Because of the biases in current data for those tested for COVID19, any study based on such data that fails to explicitly take account of the causal structure imposed by colliders and confounders will be flawed. This brings into possible dispute every conclusion made about COVID19 risk factors; those which have been claimed to be beneficial may in fact be harmful and those that are claimed to be harmful may in fact be beneficial.

While causal BNs have been used to highlight the problem, such causal graphical models also provide the solution to it. Any study based on observational data must start with consideration of the causal structure. This will show exactly what data are required; so, for example, we will need to know what proportion of people with and without some risk factor have been tested not just what proportion of those with the risk factor who were tested have COVID19. The observational data will provide some (but not necessarily all) of the probability parameters required to populate the BN model. Most of the missing parameters may be found in related studies, while others may rely on expert judgement. Once all the parameters are provided, we can use the BN inference algorithm to automatically compute unbiased conclusions. This extends to simulating 'interventions' (without the need for randomized controlled trials or new data) by 'breaking the link' from confounder variables to outcome variables.

It is important to note that we are not suggesting that studies should be based on people who have not been formally tested (unless we are interested only in whether people get COVID19 symptoms). But in the absence of sufficient random testing data, the options are either to:

- carry on using simple statistical analysis that produces misleading or flawed conclusion
- explicitly incorporate causal structure, as proposed in this paper, to take account of colliders and confounders.

Unless those undertaking statistical analysis of COVID19 data join the 'causal revolution' promoted by Pearl and others, politicians and decision-makers will continue to be fed conclusions from statistical analysis that lack validity and may be fundamentally flawed.

## Acknowledgement

This work was supported in part by the EPSRC under project EP/P009964/1: PAMBAYESIAN: Patient Managed decision-support using Bayesian Networks and The Alan Turing Institute under the EPSRC grant EP/N510129/1. Thanks to Dave Lagnado for alerting me to the Griffith et al paper and for providing valuable suggestions for improvement.